\documentclass[twocolumn,aps,showpacs,pre,amsmath]{revtex4}

\usepackage{amssymb}
\usepackage{graphicx}

\begin{document}

\title{Enhanced thermoelectric figure of merit in vertical graphene junctions}

\author{V. Hung Nguyen$^{1,2}$\footnote{E-mail: hung@iop.vast.ac.vn}, M. Chung Nguyen$^{1,2}$, Huy-Viet Nguyen$^2$, J. Saint-Martin$^1$, and P. Dollfus$^1$} \address{$^1$Institut d'Electronique Fondamentale, UMR8622, CNRS, Universit$\acute{e}$ Paris Sud, 91405 Orsay, France \\ $^2$Center for Computational Physics, Institute of Physics, Vietnam Academy of Science and Technology, P.O. Box 429 Bo Ho, 10000 Hanoi, Vietnam}

\begin{abstract}
 In this work, we investigate thermoelectric properties of junctions consisting of two partially overlapped graphene sheets coupled to each other in the cross-plane direction. It is shown that because of the weak van-der Waals interactions between graphene layers, the phonon conductance in these junctions is strongly reduced, compared to that of single graphene layer structures, while their electrical performance is weakly affected. By exploiting this effect, we demonstrate that the thermoelectric figure of merit can reach values higher than 1 at room temperature in junctions made of gapped graphene materials, for instance, graphene nanoribbons and graphene nanomeshes. The dependence of thermoelectric properties on the junction length is also discussed. This theoretical study hence suggests an efficient way to enhance thermoelectric efficiency of graphene devices.
\end{abstract}

\pacs{xx.xx.xx, yy.yy.yy, zz.zz.zz}
\maketitle

The thermoelectric effect enables direct conversion of a temperature difference into an electric voltage and vice versa, and provides a viable route for electrical power generation from waste heat. The efficiency of thermoelectric conversion is determined by the dimensionless figure of merit, $ZT$, which is given by
\begin{eqnarray}
ZT = \frac{{{G_e}{S^2}}}{{{\kappa _e} + {\kappa _p}}} T
\end{eqnarray}
where $G_e$ is the electrical conductance, $S$ is the Seebeck coefficient, and $\kappa_{e,p}$ is the thermal conductance contributed by charged carriers and lattice vibrations (phonons), respectively. For conventional materials, these transport coefficients are not independent and it is usually difficult to greatly improve their thermoelectric performance. In principle, to achieve a high $ZT$, it is simultaneously needed to suppress thermal conductance while keeping $G_e$ and $S$ less affected. Some efficient approaches \cite{dres07,snyd08} have been suggested to guide thermoelectrics studies. They are mainly based on the use of low dimensional materials and/or nanostructuring as, for instance, thin films \cite{venk01}, quantum dot supperlattices \cite{harm02}, and silicon nanowires \cite{bouk08,hoch08}.

Graphene, a 2D mono-layer material, is expected to become one of the next generation electronic materials because of its outstanding properties such as high electron mobility \cite{novo04} and high thermal conductivity \cite{sait07,seol10}. Interestingly, the two above-mentioned approaches can be naturally combined in graphene nanostructures for better thermoelectric applications. For achieving large $ZT$ in graphene systems, two important disadvantages have to be overcome: (i) $S$ is too small due to the gapless character of graphene and (ii) $\kappa_p$ is too high. Many studies to improve thermoelectric properties of graphene with different strategies of nanostructuring have been suggested. In particular, it has been shown that the Seebeck effect can be significantly enhanced in graphene nanostructures having finite energy gaps such as graphene armchair nanoribbons (GNRs) \cite{yoko13}, graphene nano-hole (nanomesh, i.e. GNM) lattices \cite{kara11}, hybrid graphene/boron nitride structures \cite{yang12}, graphene nanoribbons with a nanopore array \cite{chan12}, graphene nanoribbons consisting of alternate zigzag and armchair sections \cite{mazz11}, and graphene nanoribbons with a chevron-type geometry \cite{sevi13}. A Seebeck coefficient higher than 1 mV/K can be easily achieved in these systems, compared to values of a few tens $\mu$V/K in pristine graphene \cite{zuev09}. To suppress the thermal conductance, nano-structuring techniques such as edge roughness \cite{sevi10,kara12}, defect engineering \cite{hask11,sevi11}, isotope engineering \cite{ming10,chen12}, nano-hole lattices \cite{kara11,chan12}, hybrid heterostructures of graphene and boron nitride  \cite{yang12}, or complex geometries as in \cite{mazz11,sevi13} have been investigated. It has been shown that the thermoelectric efficiency is significantly improved in these structures and \emph{ZT} values higher than 1 can be achieved. However, there may be some issues in the use of these systems. For instance, in the GNRs with edge roughness or defects a long channel is required to achieve a low thermal conductance, which raises a question about the detrimental effects of disorder and electron-phonon scattering on their electrical performance. In the complex systems as studied in \cite{mazz11,sevi13,yang12}, the well-controlled fabrication of nanostructures may be an issue. The use of nano-hole lattices can improve the thermoelectric performance of GNRs \cite{chan12} but the improvement is still limited in 2D systems \cite{kara11}.
\begin{figure}[!t]
\centering
\includegraphics[width=3.2in]{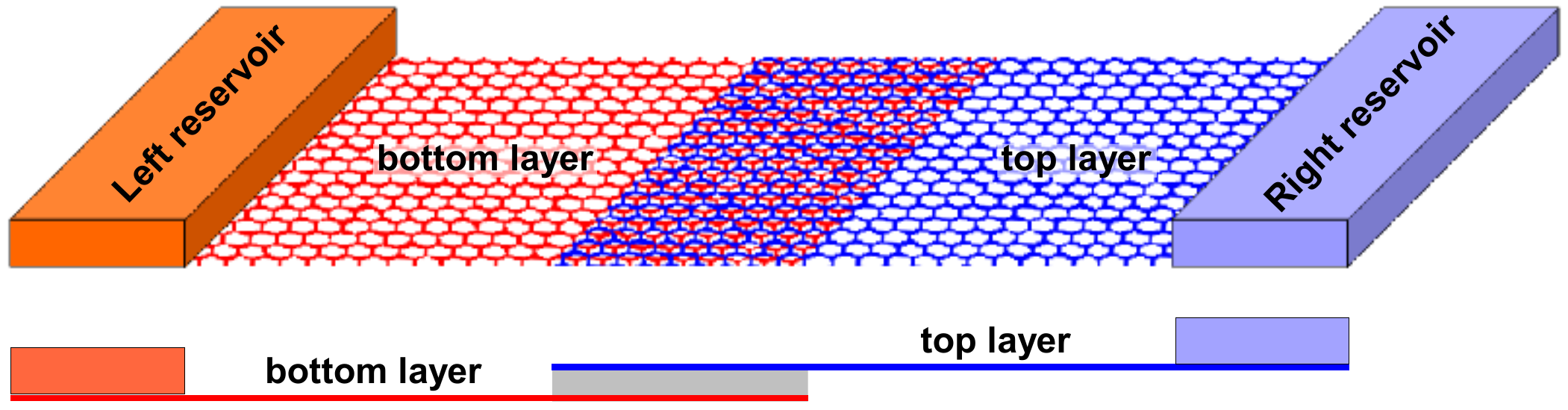}
\caption{Schematic of vertical graphene junctions investigated in this work (top) and their side view (bottom).}
\label{fig_sim1}
\end{figure}
\begin{figure*}[!t]
\centering
\includegraphics[width=5.8in]{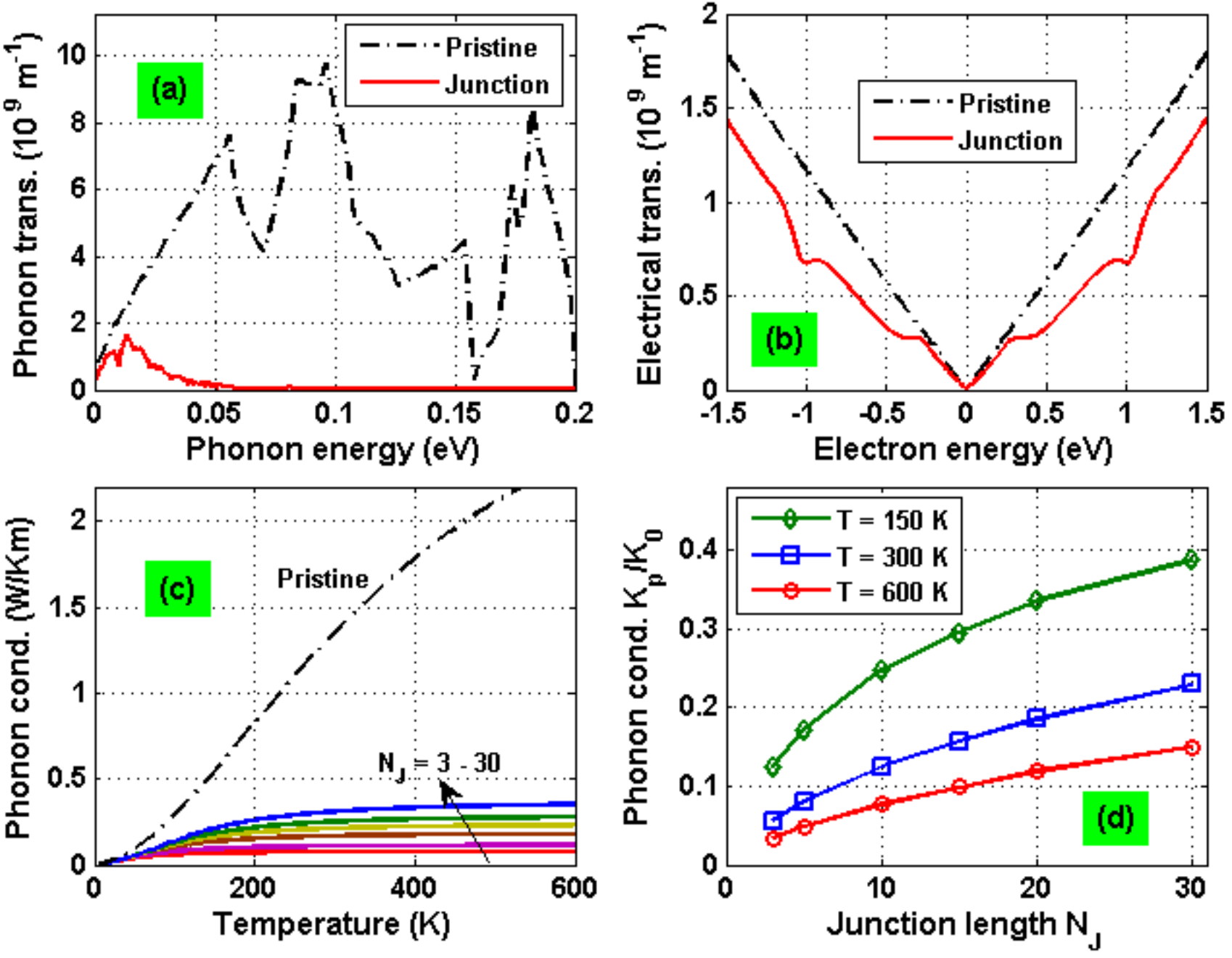}
\caption{(a) Phonon transmission and (b) electrical transmission in pristine graphene and vertical junction of two 2D graphene sheets. (c) and (d) show the phonon conductance in vertical graphene junctions as a function of temperature and junction length $N_J$, respectively, compared to the case of pristine graphene. $N_J = 5$ in (a) and (b) while $K_0$ in (d) is the phonon conductance in pristine graphene.}
\label{fig_sim4}
\end{figure*}

It is well known that the thermal properties of graphite and graphene are highly anisotropic \cite{bala11,epob12}. While the thermal conductance along the in-plane direction is high, it is hundreds of times weaker in the cross-plane direction (for graphite) \cite{cyho74}, limited by weak van-der Waals interactions between layers. This suggests an alternative way to efficiently reduce the thermal conductance in nanostructures based on multilayer graphene. In this work, we hence investigate the thermoelectric properties of graphene junctions consisting of two monolayer graphene sheets coupled to each other along the cross-plane direction, as schematized in Fig. 1. The two graphene sheets are connected to two different reservoirs. The study is motivated by the idea that as in graphite, heat flowing along the cross-plane direction in the bilayer section is weak while it is expected that the electronic conductance of the junction is still good.
\begin{figure*}[!t]
\centering
\includegraphics[width=5.8in]{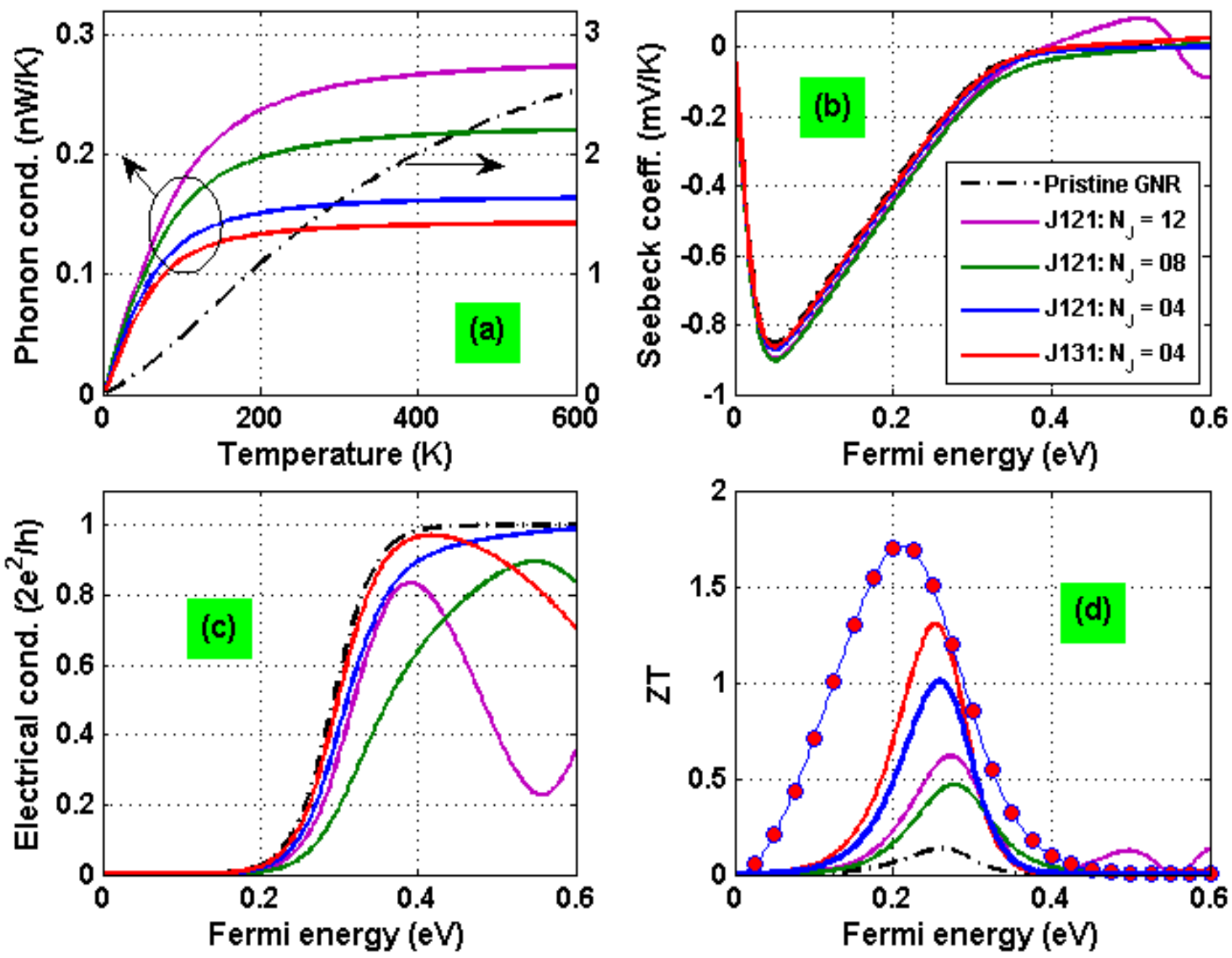}
\caption{Thermoelectric effects in vertical GNR junctions (\emph{J121}) with different junction lengths $N_J$. The junctions are made of 12-AGNRs. The data obtained in a vertical GNR junction with one more GNR layer inserted between the top and bottom layers (\emph{J131}-junction) is also presented. The quantities $G_e$, $S$ and \emph{ZT} are computed at room temperature except the solid-circle curve in (d) obtained at $T = 600$ $K$ and in the \emph{J121}-junction with $N_J = 4$. The data obtained in the pristine GNR structure (dashed-line) are plotted for comparison.}
\label{fig_sim4}
\end{figure*}

For the investigation of charge transport in the proposed system with layers arranged in Bernal stacking, calculations based on an atomistic tight-binding Hamiltonian as in \cite{hung09} were performed. The relaxation of armchair edges \cite{wson06} was also taken into account in the cases of armchair GNR. For the in-plane transport of phonons, a force constant model including four nearest neighbors \cite{kara11} was used. To describe the coupling between graphene layers, we employed the spherically symmetric interatomic potential and hence the force constant matrix elements are given by
$\Phi _{pq}^{ij} = -\alpha\exp \left( { - \beta r} \right){{r_p }{r_q }}/{r^2}$
with $\alpha = 573.76$ $N/m$ and $\beta = 20$ $nm^{-1}$ \cite{nika14}. It is worth noting that this potential model was proved to reproduce phonon dispersion in multi-layer graphene better than the familiar Lennard-Jones potential \cite{lind11}. Due to the very long mean free paths of electrons (approaching 1 $\mu m$ \cite{bolo08,mayo11}) and of phonons ($\sim$ 775 $nm$ \cite{ghos08}) at room temperature in graphene, which is much longer than the typical size of junctions under study (i.e., of about a few to few tens of $nm$), both electrons and phonons were assumed to travel ballistically and independently. The ballistic electron and phonon transmission functions were hence calculated according to the nonequilibrium Green's function formalism \cite{mazz12}. From these transmission functions, the transport quantities as electrical conductance $G_e$, Seebeck coefficient $S$ and thermal conductance $\kappa$ were computed in the Landauer formalism. In the structures under study (see Fig. 1), the key parameter is the junction length, i.e., the length of overlap zone between two graphene sheets (bilayer section), which is determined by the number of unit cells $N_J$ in this zone. Our study mainly focuses on thermoelectric effects at room temperature. Some results at 600 $K$ will be also provided to highlight the beneficial properties of the considered junctions to achieve high $ZT$ at high temperature, compared to the case of single layer graphene.

First, we present in Fig. 2 a picture showing basic transport properties of vertical junctions made of two 2D graphene sheets (\emph{J121}-junctions). Indeed, as expected, it is shown that the phonon transmission in the \emph{J121}-junction is dramatically reduced (see Fig. 2(a)), compared to the case of pristine graphene sheet, while the electrical transmission is not strongly affected (see Fig. 2(b)). In Fig. 2(c, d), we display the phonon conductance as a function of temperature and junction length $N_J$, respectively. These figures show a significant increase of phonon conductance in the \emph{J121}-junctions as a function of the junction length $N_J$. However, compared to the case of single layer graphene, only a weak increase of phonon conductance is observed as a function of temperature, especially in high temperature region \cite{scat11}. The former is essentially explained by the enhancement of interlayer coupling between two graphene sheets when increasing their overlap while the latter occurs because the contribution of phonon modes at high energies in the \emph{J121}-junctions is strongly suppressed (see Fig. 2(a)). On this basis, it is expected that a high \emph{ZT} can be easily achieved when $N_J$ is short enough in devices based on the considered two-layer junctions and at high temperature. For instance, as shown in Fig. 2(d), the ballistic phonon conductance at room temperature is reduced by $\sim$ 92 $\%$ in the case of $N_J = 5$ ($\sim$ 2.2 nm), compared to that in pristine graphene, while the electronical transmission, as previously mentioned, is not strongly affected by the use of vertical junctions.
\begin{figure*}[!t]
\centering
\includegraphics[width=5.8in]{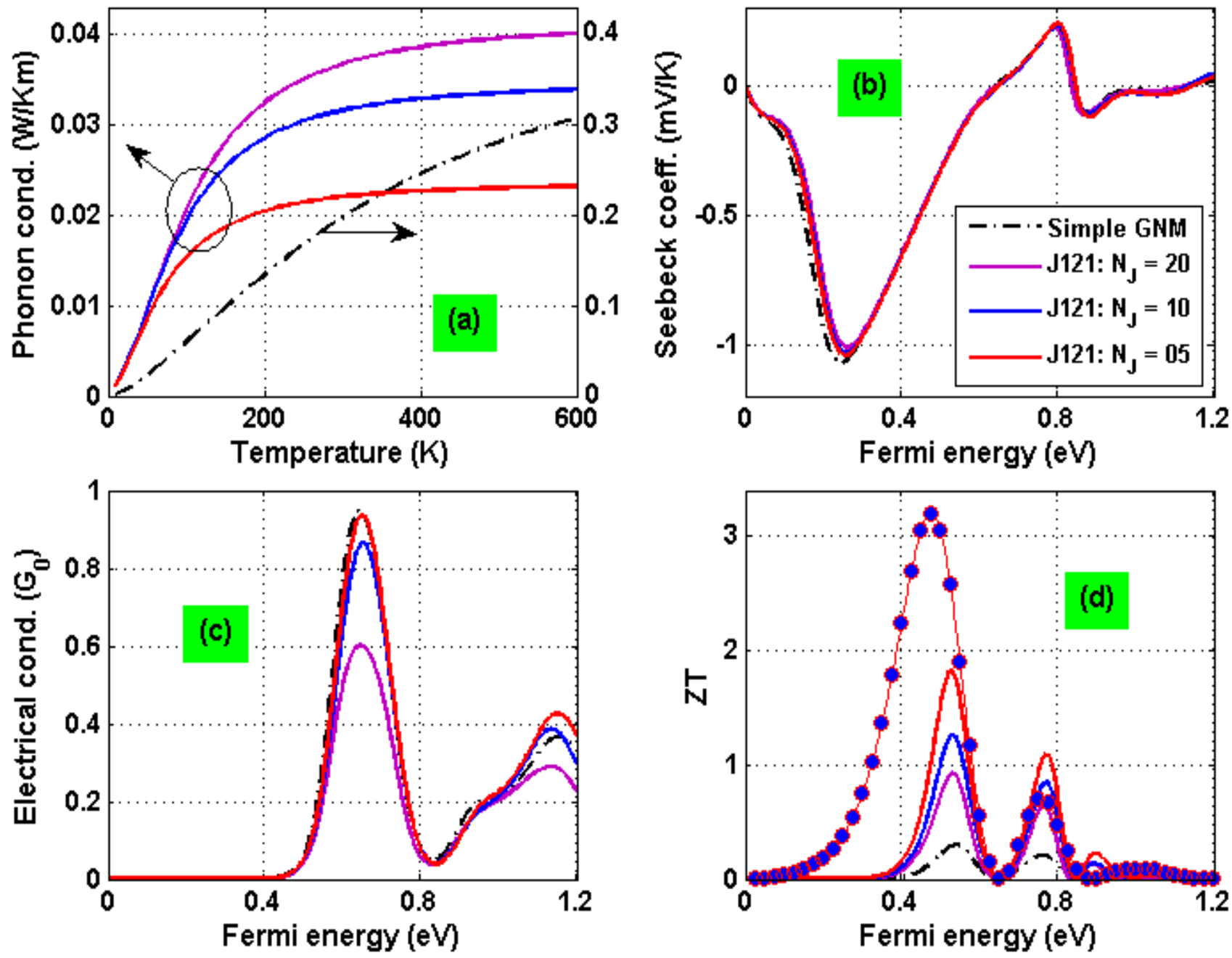}
\caption{Thermoelectric effects in vertical junctions based on GNM lattices with different junction lengths $N_J$. The GNM lattice is created in the bottom sheet. The quantities $G_e$, $S$ and \emph{ZT} are computed at room temperature except the solid-circle curve in (d) obtained at $T = 600$ $K$ and for $N_J = 5$. The data obtained in the simple GNM structure (dashed-line) is plotted for comparison. $G_0 \equiv e^2W/hL_y$.}
\label{fig_sim4}
\end{figure*}

However, the properties observed above are still not enough to achieve a strong thermoelectric effect, which in general occurs in materials having a finite energy gap (thus a high Seebeck coefficient). Therefore, we propose to replace the 2D graphene sheets in the junction by semiconducting armchair GNRs, i.e., to combine low thermal conductance and high Seebeck effect. In Fig. 3, the phonon conductance, electrical conductance, Seebeck coefficient and $ZT$ obtained in the \emph{J121}-junctions made of 12-armchair GNRs (GNRs with 12 dimers along their width \cite{mazz11}) are displayed. Here, we would like to notice that because of different electronic structures of monolayer and bilayer sections, there are confinement effects on the electron transport through the considered junctions, which results in the peaks and valleys of conductance and hence have important effects on other quantities as shown, especially for $N_J = 12$. More important, we obtained in these junctions (i) low phonon conductance (an order of magnitude smaller than that of a single layer GNR); (ii) high Seebeck coefficient (up to $\sim$ 0.9 \emph{mV/K}) and slightly degraded electrical-conductance, compared to the data obtained in the pristine GNR counterpart. As a consequence, values of $ZT$ higher than 1 can be achieved at room temperature when the junction length is short enough (see the case of $N_J = 4$ displayed in Fig. 3(d)), while the maximum value of $ZT$ is just about 0.13 in the pristine GNR. We notice once again that the phonon conductance increases and hence $ZT$ is generally reduced when increasing $N_J$. Interestingly, because of the weak increase of ballistic phonon conductance at high temperature, $ZT$ can more easily reach values higher than 1 when increasing the temperature \cite{scat11}, e.g., the maximum value of $ZT$ is about 1.72 at 600 $K$ while it is $1.02$ at 300 $K$ for the data shown in Fig. 3(d). A similar increase of $ZT$ with the temperature is also observed in the pristine GNR (not shown) but to a much smaller extent (i.e., from 0.13 at 300 $K$ to about 0.16 at 600 $K$) than in the vertical junctions.

Our study also suggests that one can further suppress the phonon conductance by inserting a few additional graphene layers between the two graphene sheets of Fig. 1, leading, therefore, to a higher $ZT$. Indeed, as displayed in Fig. 3, the data obtained in a vertical junction with one more GNR layer inserted in the overlap region (\emph{J131}-junction) shows a reduced phonon conductance and an enhanced $ZT$ ($\sim 1.31$ at room temperature), compared to the case of \emph{J121}-junctions ($\sim 1.02$). However, we also find that inserting more graphene layers into the junction is not really an efficient way to further enhance $ZT$. It is because, while the reduction of phonon conductance is weak, the electronic transmission of the junction can be significantly degraded by increasing the number of graphene layers. Thus, the $J121$- and $J131$-junctions are probably the best compromise to get the balance between high electrical performance and low thermal conductance.

We now discuss the way to achieve high $ZT$ in 2D graphene channels. To have a high thermoelectric effect in a 2D graphene system, we propose to create a nano-hole lattice in the bottom graphene sheet, which opens a finite energy gap \cite{kara11,lian10,hung13}. Here, we have considered a square graphene nanomesh (GNM) lattice with periodic nanoholes obtained by removal of 26 carbon atoms in a super-cell of 140 atoms. Note that the use of GNM lattices in the top or in both graphene sheets can have the same effects. We display in Fig. 4 the phonon conductance, electrical conductance, Seebeck coefficient and $ZT$ obtained in the GNM-based $J121$ junctions. The occurrence of conductance peaks and valleys is essentially due to mini-bands and mini-gaps in the bandstructure of GNM lattices, e.g., as shown in \cite{kara11}. In the case of a simple GNM lattice, even if $S$ is enhanced, $ZT$ is limited \cite{kara11} (e.g., its maximum value is about 0.3 here) because the phonon conductance is still high. Obviously, the thermoelectric effect is greatly improved in the GNM-based $J121$ junctions. Again, while the electrical performance ($G_e$ and $S$) is weakly affected (see Fig. 4(b,c)), the phonon conductance is strongly reduced, i.e., by a factor greater than ten at high temperature and with a short junction (see Fig. 4(a)), compared to the case of the simple GNM lattice. Therefore, $ZT$ can easily reach values higher than 1 at room temperature for $N_J \leq 10$ (see Fig. 4(d)). It can even get a higher value of $\sim$ 3.21 at 600 $K$ \cite{scat11}.

In conclusion, by exploiting the weak van-der Waals interactions between graphene layers, we have shown that the use of junctions consisting of two partially overlapped graphene layers can enhance significantly the thermoelectric efficiency of graphene devices. This type of junctions allows for greatly reducing the phonon conductance while keeping the electronic performance less affected. We have demonstrated that using this strategy, values of $ZT$ higher than 1 can be easily achieved in junctions made of armchair GNRs or GNM lattices. The study also suggests that by combining this type of junctions with other techniques of bandgap engineering, one can achieve very high values of $ZT$ in graphene devices.

\textbf{\textit{Acknowledgment.}} This research in Hanoi is funded by Vietnam's National Foundation for Science and Technology Development (NAFOSTED) under grant number 103.02-1012.42.

\end{document}